\documentclass[iop]{emulateapj}
\usepackage{amsmath, lettrine}
\usepackage{placeins}
\usepackage{booktabs}
\usepackage{color}
\usepackage{cleveref, array}
\usepackage{subfigure}
\usepackage{graphicx}
\graphicspath{{./HydroThickThin/}}

\shorttitle{The Thomson scattering cross section in a magnetized, high density plasma}
\shortauthors{A.F.A. Bott and G. Gregori}
\begin{document}
\newcolumntype{C}[1]{>{\centering\arraybackslash$}m{#1}<{$}}
\newlength{\mycolwd}                                         
\settowidth{\mycolwd}{$s + \chi_0\gamma k^2 +(4+\beta){\kappa_P \over C_{V0}}\rho T_0^3$}

\title{The Thomson scattering cross section in a magnetized, high density plasma}



\author{A. F. A. Bott\altaffilmark{1,*} and G. Gregori\altaffilmark{1}}

\altaffiltext{1}{Clarendon Laboratory, University of Oxford, Parks Road, Oxford OX1 3PU, United Kingdom}
\altaffiltext{*}{archie.bott@physics.ox.ac.uk}



\begin{abstract}
We calculate the Thomson scattering cross section in a non-relativistic, magnetized, high density plasma -- in a regime where collective excitations can be described by magnetohydrodynamics. We show that, in addition to cyclotron resonances and an elastic peak, the cross section exhibits two pairs of peaks associated with slow and fast magnetosonic waves; by contrast, the cross section arising in pure hydrodynamics possesses just a single pair of Brillouin peaks.   
Both the position and the width of these magnetosonic-wave peaks depend on the ambient magnetic field and temperature, as well as transport and thermodynamic coefficients, and so can therefore serve as a diagnostic tool for plasma properties that are otherwise challenging to measure. \\


\end{abstract}

\section{Introduction}
Understanding radiation transport, opacity and thermodynamic properties of strongly coupled, magnetized plasmas is important for 
modelling the atmosphere of magnetars and neutron stars \citep{meszaros},
dynamo formation and evolution in planetary and stellar interiors \citep{kulsrud,guillot}, as well as inertial confinement fusion \citep{lindl,remington}.
An important quantity that determines the opacity of these plasmas is the Thomson scattering cross section \citep{meszaros,crowley}. Moreover, while Thomson scattering of laser light is used as a plasma diagnostic tool \citep{evans}, understanding the measurements in presence of a background magnetic field has, so far, been limited to special cases of negligible correlations between the electrons \citep{herold,meszaros}, or weakly coupled plasmas at wavelengths below the mean free path of constituent particles~\citep{salpeter,froula}.


In this paper we will calculate the non-relativistic Thomson scattering cross section associated with collective excitations of a magnetized, strongly coupled plasma -- that is, a plasma where the motion of charged particles is determined by both the presence of an ambient magnetic field, ${\bf B}_0$, and by short and long range correlations between all the particles in the system. Such excitations are most appropriately described by magnetohydrodynamics (MHD).

The structure of this paper is as follows. 
The double-differential Thomson scattering cross section is presented in Section \S\ref{cross_section}, and its relationship to the dynamic structure factor discussed.
In Section \S\ref{GenEqMHD}, we write 
down the governing equations of MHD in a standard form, and then derive 
evolution equations in terms of density, bulk fluid velocity, magnetic field and 
temperature, as well as constitutive parameters of the matter.  Section \S\ref{Fluctuations} 
provides a derivation of the density autocorrelation function -- and thereby the dynamic structure factor 
-- arising in MHD for small-amplitude fluctuations. Before considering the 
general case in Section \S\ref{Fluctuations_oblique}, we focus on the special 
cases of fluctuations whose wavevector is parallel to the magnetic field (Section 
\S\ref{Fluctuations_par}), and quasi-perpendicular fluctuations 
(Section \S\ref{Fluctuations_perp}). Considering these particular cases -- which can be treated analytically --
allows for the clearest physical interpretation of the characteristic scattering peaks emerging in the MHD model for general fluctuations.  
Section \S\ref{Fluctuations_oblique} also considers the case of fluctuations in 
dense plasmas where the magnetic energy density is only a finite fraction of 
the thermal energy density; this in turn anticipates how the dynamic structure 
factor might be altered from a purely hydrodynamic picture at sufficiently large 
magnetic field strengths.
Finally, in Section \S\ref{balance} we briefly discuss how to extend our model to include quantum effects.

\section{Thomson scattering cross section}
\label{cross_section}
Even neglecting particle correlations, calculating the scattering cross sections of photons by single electrons is a non-trivial task when there is a background magnetic field. This is because the motion of the electron under the influence of the incident electric field is altered by the presence of the background, introducing resonances at the cyclotron frequency \citep{meszaros}; thus the full polarization tensor, $\pmb{\mathcal{P}}$, must be accounted for. Assuming that the ambient magnetic field is smaller than the Schwinger's field (thus neglecting vacuum polarization effects), we have that the differential cross section $d\sigma$ for Thomson scattering from a single electron into a solid angle $d\Omega$ is
\begin{equation}
    \frac{d \sigma}{d \Omega} = r_e^2 
    |\langle \hat{\boldsymbol{e}}^{(1)}| \pmb{\mathcal{P}} |\hat{\boldsymbol{e}}^{(0)} \rangle |^2, \label{TScrosssec_mag_gen} 
\end{equation}
where $r_e=e^2/4\pi \epsilon_0 m_e c^2$ is the classical electron radius, and $\hat{\boldsymbol{e}}^{(0)}$ and $\hat{\boldsymbol{e}}^{(1)}$ are the incident and scattered photon polarization, respectively. 
Equation (\ref{TScrosssec_mag_gen}) can be simplified somewhat by noting that the polarization matrix is diagonal in the rotated frame where 
$\hat{\boldsymbol{e}}^{(0)} \equiv \left[ {\boldsymbol{e}}_+^{(0)}, {\boldsymbol{e}}_-^{(0)}, {\boldsymbol{e}}_z^{(0)} \right]$, with ${\boldsymbol{e}}_z^{(0)}$ along the direction of ${\bf B}_0$ and ${\boldsymbol{e}}_{\pm}^{(0)}={\boldsymbol{e}}_x^{(0)}\pm i {\boldsymbol{e}}_y^{(0)}$. A similar decomposition applies to the scattered photon polarization. 

A further simplification applies if the electron temperature is much smaller that its rest mass energy (i.e., in the non-relativistic regime). In this case, the polarization matrix is independent of the electron velocity, and
we thus have \citep{herold,ventura,nagel}
\begin{eqnarray}
    \frac{d \sigma}{d \Omega} &=& r_e^2 \left|
    \frac{\langle {\boldsymbol{e}}^{(1)}_+| {\boldsymbol{e}}^{(0)}_+ \rangle}{1+\xi^{1/2}+i\gamma_R} +
    \frac{\langle {\boldsymbol{e}}^{(1)}_-| {\boldsymbol{e}}^{(0)}_- \rangle}{1-\xi^{1/2}+i\gamma_R} +
    \frac{\langle {\boldsymbol{e}}^{(1)}_z| {\boldsymbol{e}}^{(0)}_z\rangle}{1+i\gamma_R}
    \right|^2 \nonumber \\
    &\equiv& r_e^2 |f_{(\hat{\boldsymbol{e}}^{(1)},\hat{\boldsymbol{e}}^{(0)})}|^2,
    \label{cross}
\end{eqnarray}
where $\xi=(\omega_{ce}/\omega_0)^2$, with $\omega_0$ the incident photon frequency, $\omega_{ce}=e B_0/m_e$ the electron cyclotron frequency, and
$\gamma_R \omega_{ce}^2/\omega_0 = e^2 \omega_{ce}^2/6 \pi \epsilon_0 m_e c^3 \approx 4\times 10^{15} (B_0/10^{12}\,{\rm G})^2$ s$^{-1}$ is the radiative damping coefficient. 
While the above expression uses a value for the background magnetic field that is typical for magnetars, the actual damping coefficient $\gamma_R$ is independent of $B_0$. Equation (\ref{cross}) applies for $\omega_p^2 \ll \omega_0^2$, where $\omega_p$ is the plasma frequency.
We see that the cross section is strongly enhanced at the cyclotron resonance, corresponding to the resonant photon absorption and re-emission between Landau levels. 
This is an effect that is not 
present when there is no ambient magnetic field.  

Since in the non-relativistic limit 
the cross section is the same for all electrons, the quantity $f_{(\hat{\boldsymbol{e}}^{(1)},\hat{\boldsymbol{e}}^{(0)})}$ plays the same role as a scattering form factor. We now consider a system consisting of many electrons in a plasma which is strongly coupled. Because the plasma can sustain different type of waves, energy can be exchanged between the incident photons and the waves. The double-differential cross section then reads as \citep{crowley}
\begin{equation}
   \frac{d^2 \sigma}{d \Omega d \omega_1} = N_e r_e^2 |f_{(\hat{\boldsymbol{e}}^{(1)},\hat{\boldsymbol{e}}^{(0)})}|^2 \frac{\omega_1}{\omega_0} S_{ee}(\boldsymbol{k},\omega),
\end{equation}
where $N_e$ is the total number of electrons, $\omega_1$ the scattered photon frequency, $\omega=\omega_0-\omega_1$ and
${\boldsymbol{k}}={\boldsymbol{k}}_0-{\boldsymbol{k}}_1$ (for ${\boldsymbol{k}}_0$ the incident wavevector, and ${\boldsymbol{k}}_1$ the scattered wavevector). 
We note that the response of the system is anisotropic with respect to the direction of the background magnetic field. 
The quantity $S_{ee}(\boldsymbol{k},\omega)$, known as the dynamic structure factor, encodes the angular and energy distribution of the scattering that results from collective motions of the electrons.
While the above expression applies to free electrons in the plasma, it can be generalized to the case of an electron-ion plasma \citep{chihara,gregori,crowley2}. Focusing on low-frequency excitations, the relevant part of the cross section reads as
\citep{gregori,crowley2}
\begin{equation}
\frac{d^2 \sigma}{d \Omega d \omega_1} = N r_e^2 |f_{(\hat{\boldsymbol{e}}^{(1)},\hat{\boldsymbol{e}}^{(0)})}|^2 \frac{\omega_1}{\omega_0} |f_I(k) + q(k)|^2 S_{n n}(k,\omega),  
\end{equation}
where $N$ is the total number of ions, 
$f_I(k)$ is the ion form factor (accounting for the bound-electron correlations) and $q(k)$ is the screening cloud of kinematically bound free electrons that follow the ion. 
The dynamic structure factor $S_{n n}(k,\omega)$ is defined by \citep{hansen,crowley2}
\begin{equation}
\label{DSF}
S_{n n}(\boldsymbol{k} ,\omega) = {1 \over 2 \pi N}\int \mathrm{d}t \, e^{i \omega t} \langle n(\boldsymbol{k},t) n(-\boldsymbol{k},0) \rangle \, ,
\end{equation}
where $n(\boldsymbol{k},t)$ is the Fourier component of the number density with wavevector $\boldsymbol{k}$ (wavenumber $k \equiv |\boldsymbol{k}|$), and fluctuation frequency 
$\omega$. The operator $\langle ... \rangle$ corresponds to a thermal average over the particles' ensemble. 

It is clear that knowledge of the dynamic structure factor is essential for a detailed derivation of the scattering spectrum incorporating collective excitations. For magnetized, strongly coupled plasma, the model
of those excitations we choose to employ to calculate the dynamic structure factor is that of magnetohydrodynamics (MHD), accounting for viscosity, electrical resistivity and 
heat conductivity. A fluid model of this sort is suitable for plasma whose characteristic mean-free path $\lambda_{\rm mfp}$ of constituent particles due to Coulomb collisions is much smaller than the typical length scales $k^{-1}$ associated with the fluctuations of interest. In a strongly coupled plasma, for which $\lambda_{\rm mfp} \lesssim \lambda_{\rm De}$ (where $\lambda_{\rm De}$ is the plasma Debye length), it follows that collective excitations -- that is, excitations for which $k \lambda_{\rm De}  \ll 1$ -- must satisfy $k \lambda_{\rm mfp} \ll 1$. We conclude that MHD is the appropriate model for calculating the dynamic structure factor for magnetized, strongly coupled plasma.


For completeness, we observe that at high densities and with magnetic fields significantly below the critical value, Coulomb collisions can also alter the Thomson scattering cross section associated with non-collective excitations in strongly coupled plasma  from that of classical weakly coupled plasma. More specifically, the de-excitation rate of the Landau levels, and the radiative damping coefficient must be changed \citep{nagel} to
$ \gamma_R \rightarrow \gamma_R+\gamma_{coll}$, where
\begin{equation}
\frac{\gamma_{coll} \omega_{ce}^2}{\omega_0}=3.1\times 10^8 \left(\frac{B_0}{10^{12}\,{\rm G}}\right)^{-3/2} \left(\frac{n_e}{10^{20}\,{\rm cm^{-3}}}\right) Z^2\,\,\rm s^{-1},
\end{equation}
with $Z$ the ion charge. Thus, in dense plasmas, collisional processes can significantly broaden the cyclotron resonances. 




\section{General equations for magnetohydrodynamics (MHD)}
\label{GenEqMHD}
We now focus on the calculation of $S_{n n}(k,\omega)$ in a magnetized plasma. The approach we use begins by writing down the relevant fluid equations and then applying a linearization procedure in order to derive the density fluctuations \citep{hansen2,schmidt,cross}.
The governing equations of MHD are conservation laws of mass, momentum, magnetic 
flux and internal energy:
\begin{subequations}
\begin{eqnarray}
\label{MHDequations}
\frac{\mathrm{d} \rho}{\mathrm{d} t}  & = & - \rho \nabla \cdot \boldsymbol{u}, 
 \label{MHDgoveqns_mass} \\
\rho {\mathrm{d} \boldsymbol{u} \over \mathrm{d} t} & = & -\nabla p - \nabla \cdot \boldsymbol{\Pi} + \frac{\left(\nabla \times \boldsymbol{B}\right) \times \boldsymbol{B}}{\mu_0},  \label{MHDgoveqns_mom} \\ 
 {\mathrm{d} \boldsymbol{B} \over \mathrm{d} t}  & = & \boldsymbol{B} \cdot \nabla \boldsymbol{u} - \boldsymbol{B} \nabla \cdot \boldsymbol{u} - \nabla \times \left(\eta \nabla \times \boldsymbol{B} \right),  \label{MHDgoveqns_flux} \\
\rho {\mathrm{d} \epsilon \over \mathrm{d} t} & = & -p \nabla \cdot \boldsymbol{u} - \boldsymbol{\Pi}:\nabla \boldsymbol{u} + \eta \frac{\left|\nabla \times \boldsymbol{B}\right|^2}{\mu_0} - \nabla \cdot \boldsymbol{q} , \qquad    \label{MHDgoveqns_eng}
\end{eqnarray} 
\end{subequations}
where $\rho=M n$ is the mass density (with $M$ the ion mass), $t$ is time, $\boldsymbol{u}$ the bulk fluid velocity,
\begin{equation}
 \frac{\mathrm{d} }{\mathrm{d} t}  \equiv \frac{\partial }{\partial  t} + 
 \boldsymbol{u} \cdot \nabla 
\end{equation}
the convective derivative, $p$ the pressure, $\boldsymbol{\Pi}$ the viscosity tensor, $\boldsymbol{B}$ the magnetic field, $\mu_0$ the permeability of free space, $\eta$ the (assumed isotropic) resistivity, $\epsilon$ the internal energy, and $\boldsymbol{q}$ is the heat flux.  

For subsequent calculations, it is helpful to rewrite the internal energy conservation law (\ref{MHDgoveqns_eng}) 
as an evolution equation for the fluid temperature $T$ in terms of density, bulk flow velocity and the magnetic field. We do this by using the 
first law of thermodynamics, 
\begin{equation}
 \frac{\mathrm{d} \epsilon}{\mathrm{d} t}   = T  \frac{\mathrm{d} S}{\mathrm{d} t} 
 + \frac{p}{\rho^2}  \frac{\mathrm{d} \rho}{\mathrm{d} t} ,
\end{equation}
where $S$ is the specific entropy, to write down a conservation law for 
specific entropy:
\begin{equation}
\rho T {\mathrm{d} S \over \mathrm{d} t} = -\boldsymbol{\Pi}:\nabla \boldsymbol{u} + \eta \frac{\left|\nabla \times \boldsymbol{B}\right|^2}{\mu_0} - \nabla \cdot \boldsymbol{q} 
 . \label{MHDgoveqns_ent}
\end{equation}
In turn, it can be shown using thermodynamic identities (see Appendix \ref{AppendixThermo}) that
\begin{equation}
 {\mathrm{d} S \over \mathrm{d} t} = \frac{C_V}{T}  \left({\mathrm{d} T \over \mathrm{d} t} 
 - \frac{\gamma - 1}{\alpha_T} {\mathrm{d} \rho \over \mathrm{d} t} \right), \label{thermoiden_A}
\end{equation}
\color{black} 
where $C_V$ is the heat capacity at constant volume, $\gamma$ the adiabatic 
index, and $\alpha_T$ the coefficient of thermal expansion. Thus, we deduce that the temperature evolves  
according to
\begin{eqnarray}
\rho C_V {\mathrm{d} T \over \mathrm{d} t} & = & - \frac{\gamma - 1}{\alpha_T} \rho C_V \nabla \cdot \boldsymbol{u} - \boldsymbol{\Pi}:\nabla \boldsymbol{u} \nonumber\\ 
&& \quad + \eta \frac{\left|\nabla \times \boldsymbol{B}\right|^2}{\mu_0} - \nabla \cdot \boldsymbol{q} 
 . \label{MHDgoveqns_temp} 
\end{eqnarray}

Finally in the section, we write the governing equations (\ref{MHDgoveqns_mass}), 
(\ref{MHDgoveqns_mom}), (\ref{MHDgoveqns_flux}) and (\ref{MHDgoveqns_temp})
in terms of only density, bulk flow velocity, magnetic field and temperature, 
and constitutive parameters of the fluid -- in other words, we substitute for the 
pressure $p$, the viscosity tensor $\boldsymbol{\Pi}$ and the heat flux $\boldsymbol{q}$ in terms of the aforementioned 
variables. To eliminate the pressure, we use thermodynamic identity
\begin{equation}
\nabla p = \frac{c_s^2}{\gamma} \left(\nabla \rho + \rho \alpha_T \nabla T\right) 
, \label{thermoiden_B}
\end{equation}
for $c_s$ the adiabatic sound speed (see Appendix \ref{AppendixThermo}). For the viscosity 
tensor and heat flux, we use constitutive relations
\begin{subequations}
\label{constiteqns}
\begin{eqnarray}
\boldsymbol{\Pi} & = & -\zeta_s \left[\nabla \boldsymbol{u} + \left(\nabla  \boldsymbol{u}\right)^{\rm T} - \frac{2}{3} \left(\nabla \cdot \boldsymbol{u} \right) \mathbf{I} \right] 
- \zeta_b \left(\nabla \cdot \boldsymbol{u} \right) \mathbf{I} , \qquad \; \\
\boldsymbol{q} & = & - \kappa \nabla T ,
\end{eqnarray}
\end{subequations}
where $\zeta_s$ is the first coefficient of viscosity (or shear viscosity), $\zeta_b$ 
the second coefficient of viscosity (or bulk viscosity), $\mathbf{I}$ the 
identity tensor, and $\kappa$ the thermal conductivity. We note that for sufficiently large magnetic fields, the chosen constitutive relations may not be appropriate; for example, it is well known that $\boldsymbol{q}$ is predominantly parallel to $\boldsymbol{B}$ in weakly-coupled collisional plasma where the Larmor radius $r_{ce}$ of constituent thermal electrons satisfies $r_{ce} \ll \lambda_{\rm mfp}$~\citep{braginskii}. However, the calculation presented here is easily modified to include such effects if necessary, which in any case do not affect any of our key results qualitatively.

On substituting (\ref{constiteqns}), we find our desired system of equations:
\begin{subequations}
\begin{eqnarray}
\frac{\mathrm{d} \rho}{\mathrm{d} t}  & = & - \rho \nabla \cdot \boldsymbol{u}, 
 \label{MHDgoveqns_dens} \\
\rho {\mathrm{d} \boldsymbol{u} \over \mathrm{d} t} & = & -  \frac{c_s^2}{\gamma} \left(\nabla \rho + \rho \alpha_T \nabla T\right) \nonumber \\
&& \, + \frac{\left(\nabla \times \boldsymbol{B}\right) \times \boldsymbol{B}}{\mu_0} + \nabla \left(\zeta_b \nabla \cdot \boldsymbol{u} \right) \nonumber \\
&& \, + \nabla \cdot \left(\zeta_s \left[\nabla \boldsymbol{u} + \left(\nabla  \boldsymbol{u}\right)^{\rm T} - \frac{2}{3} \left(\nabla \cdot \boldsymbol{u} \right) \mathbf{I} \right] \right), \qquad \label{MHDgoveqns_flow} \\ 
 {\mathrm{d} \boldsymbol{B} \over \mathrm{d} t}  & = & \boldsymbol{B} \cdot \nabla \boldsymbol{u} - \boldsymbol{B} \nabla \cdot \boldsymbol{u} - \nabla \times \left(\eta \nabla \times \boldsymbol{B} \right),  \label{MHDgoveqns_fluxB} \\
\rho {\mathrm{d} T \over \mathrm{d} t} & = & - \frac{\gamma - 1}{\alpha_T} \rho \nabla \cdot \boldsymbol{u} - \frac{1}{C_V} \boldsymbol{\Pi}:\nabla \boldsymbol{u} \nonumber\\ 
&& \quad + \eta \frac{\left|\nabla \times \boldsymbol{B}\right|^2}{\mu_0 C_V} + \frac{1}{C_V} \nabla \cdot \left(\kappa \nabla T\right) .    \label{MHDgoveqns_tempB}
\end{eqnarray} 
\end{subequations}
where for brevity we have not written out in full the viscous dissipation term 
in the temperature evolution equation. Note that these equations do not necessarily assume 
that the matter's transport coefficients (specifically $\zeta_s$, $\zeta_b$, $\eta$ and $\kappa$) 
are independent of temperature or density. 

\section{Fluctuations, and the dynamic structure factor}
\label{Fluctuations}

We now evaluate the MHD dynamic structure factor in the limit of small-amplitude 
fluctuations. To perform this calculation, we consider some equilibrium state, 
with density $\rho_0$, no bulk flow motion, magnetic field 
$\boldsymbol{B}_0$, temperature $T_0$, sound speed $c_{s0}$, adiabatic index 
$\gamma_0$, coefficient of thermal expansion $\alpha_{T0}$, bulk viscosity 
$\zeta_{b0}$, shear viscosity $\zeta_{s0}$, resisivity $\eta_0$, specific heat 
capacity at constant volume $C_{V0}$, and thermal conductivity $\kappa_0$. We 
then consider small-amplitude fluctuations of dynamic quantities on this 
equilibrium: 
\begin{equation}
\rho = \rho_0 + \delta \rho, \quad  \boldsymbol{u} = \delta \boldsymbol{u}, 
\quad \boldsymbol{B} = \boldsymbol{B}_0 + \delta \boldsymbol{B}, \quad T = 
T_0 + \delta T .  \label{fluctuations_def}
\end{equation}
Substituting (\ref{fluctuations_def}) into (\ref{MHDgoveqns_dens}), 
(\ref{MHDgoveqns_flow}), (\ref{MHDgoveqns_fluxB}) and (\ref{MHDgoveqns_tempB}), 
and neglecting terms quadratic or higher in fluctuating quantities, we find
\begin{subequations}
 \label{MHDlineqns}
\begin{eqnarray}
\frac{\partial \delta \rho}{\partial  t}  & = & - \rho_0 \nabla \cdot \delta \boldsymbol{u}, 
 \label{MHDlineqns_dens} \\
\rho_0 {\partial \delta \boldsymbol{u} \over \partial  t} & = & -  \frac{c_{s0}^2}{\gamma_0} \left(\nabla \delta \rho + \rho_0 \alpha_{T0} \nabla \delta T\right) \nonumber \\
&& \, + \frac{\boldsymbol{B}_0 \cdot \nabla \delta \boldsymbol{B}}{\mu_0} -\nabla \left( \frac{\boldsymbol{B}_0 \cdot \delta \boldsymbol{B}}{\mu_0} \right) \nonumber \\
&& \, +  \zeta_{s0} \nabla^2 \delta \boldsymbol{u} +\zeta_{c0} \nabla \left(\nabla \cdot \delta \boldsymbol{u} \right) , \qquad \label{MHDlineqns_flow} \\ 
{\partial \delta \boldsymbol{B} \over \partial  t}  & = & \boldsymbol{B}_0 \cdot \nabla \delta \boldsymbol{u} - \boldsymbol{B}_0 \nabla \cdot \delta \boldsymbol{u} + \eta_0 \nabla^2 \delta \boldsymbol{B},  \label{MHDlineqns_flux} \\
{\partial \delta T \over \partial  t} & = & - \frac{\gamma_0 - 1}{\alpha_{T0}} \nabla \cdot \delta \boldsymbol{u} + \gamma_0 \chi_0 \nabla^2  \delta T .    \label{MHDlineqns_temp}
\end{eqnarray} 
\end{subequations}
where we have defined ``compressive'' viscosity coefficient $\zeta_{c0} \equiv \zeta_{b0} - 2 
\zeta_{s0}/3$, and thermal diffusivity $\chi_0 \equiv \kappa_0/\rho_0 C_{V0} 
\gamma_0$.

To find the dynamic structure factor, we transform equations (\ref{MHDlineqns}) using a 
Fourier transform in space, and a Laplace transform in time. For vector quantity 
$\delta \boldsymbol{x}$, this operation is defined as
\begin{equation*}
\widetilde{\delta \boldsymbol{x}}_{\boldsymbol{k}}(s) = \int_0^{\infty} \!\! \mathrm{d}t \, e^{-st}  \int_{- \infty}^{+ \infty} \! \! \mathrm{d}^3\boldsymbol{r}\, e^{i \boldsymbol{k} \cdot \boldsymbol{r}}\,\delta \boldsymbol{x}(\boldsymbol{r},t).
\end{equation*}
Applying this, and using standard properties of Laplace and Fourier transforms 
under derivatives, we find 
\begin{subequations}
 \label{MHDlintranseqns}
\begin{eqnarray}
s \widetilde{\delta \rho}_{\boldsymbol{k}}(s) & = & - \rho_0 i \boldsymbol{k} \cdot \widetilde{\delta \boldsymbol{u}}_{\boldsymbol{k}}(s) +  {\delta \rho}_{\boldsymbol{k}}(0) , 
 \label{MHDlintranseqns_dens} \\
\rho_0 s \widetilde{\delta \boldsymbol{u}}_{\boldsymbol{k}}(s) & = & -  \frac{c_{s0}^2}{\gamma_0} \left(i \boldsymbol{k} \widetilde{\delta \rho}_{\boldsymbol{k}}(s) + i \rho_0 \alpha_{T0} \boldsymbol{k} \widetilde{\delta T}_{\boldsymbol{k}}(s)\right) \nonumber \\
&& \, + i \widetilde{\delta \boldsymbol{B}}_{\boldsymbol{k}}(s) \frac{\boldsymbol{B}_0 \cdot \boldsymbol{k}}{\mu_0} -i \boldsymbol{k} \frac{\boldsymbol{B}_0 \cdot \widetilde{\delta \boldsymbol{B}}_{\boldsymbol{k}}(s)}{\mu_0} \nonumber \\
&& \, - \zeta_{s0} k^2 \widetilde{\delta \boldsymbol{u}}_{\boldsymbol{k}}(s) -\zeta_{c0} \boldsymbol{k} \left(\boldsymbol{k} \cdot \widetilde{\delta \boldsymbol{u}}_{\boldsymbol{k}}(s) \right) \nonumber \\
&& \quad + \rho_0 {\delta \boldsymbol{u}}_{\boldsymbol{k}}(0), \qquad \label{MHDlintranseqns_flow} \\ 
s \widetilde{\delta \boldsymbol{B}}_{\boldsymbol{k}}(s)  & = & i \left(\boldsymbol{k} \cdot \boldsymbol{B}_0 \right) \widetilde{\delta \boldsymbol{u}}_{\boldsymbol{k}}(s) - i\boldsymbol{B}_0 \left(\boldsymbol{k} \cdot \widetilde{\delta \boldsymbol{u}}_{\boldsymbol{k}}(s) \right) \nonumber \\
&& \quad - \eta_0 k^2 \widetilde{\delta \boldsymbol{B}}_{\boldsymbol{k}}(s) + {\delta \boldsymbol{B}}_{\boldsymbol{k}}(0),  \label{MHDlintranseqns_flux} \\
s \widetilde{\delta T}_{\boldsymbol{k}}(s) & = & - i \frac{\gamma_0 - 1}{\alpha_{T0}} \boldsymbol{k} \cdot \widetilde{\delta \boldsymbol{u}}_{\boldsymbol{k}}(s) - \gamma_0 \chi_0 k^2 \widetilde{\delta T}_{\boldsymbol{k}}(s) \nonumber \\
&& \quad + {\delta T}_{\boldsymbol{k}}(0).    \label{MHDlintranseqns_temp}
\end{eqnarray} 
\end{subequations}
The dynamic structure factor $S_{nn}(\boldsymbol{k},\omega)$ is related to the transformed fluctuating 
quantities via the following limit of the density autocorrelation function \citep{hansen}:
\begin{equation}
\label{DSFlimit}
{S_{nn}(\boldsymbol{k},\omega) \over S_{nn}(\boldsymbol{k})} = 2 \Re \left[ \lim_{\varepsilon \to 0}{\langle\delta \rho_{\boldsymbol{k}}^*(0)\widetilde{\delta\rho}_{\boldsymbol{k}}(s=\varepsilon +i \omega)\rangle \over \langle \delta \rho^*_{\boldsymbol{k}}(0)\delta\rho_{\boldsymbol{k}}(0)\rangle} \right],
\end{equation}
where we have used the definition
\begin{equation}
S_{nn}(\boldsymbol{k})=\int S_{nn}(\boldsymbol{k},\omega) d\omega,
\end{equation}
which is usually referred to as the static structure factor.

To find an explicit expression for $S(\boldsymbol{k},\omega)$, we solve equations (\ref{MHDlintranseqns}) 
for $\widetilde{\delta \rho}_{\boldsymbol{k}}(s)$, before evaluating the density 
autocorrelation function. We find that
\begin{equation}
\frac{\langle\delta \rho_{\boldsymbol{k}}^*(0)\widetilde{\delta\rho}_{\boldsymbol{k}}(s)\rangle}{ \langle \delta \rho^*_{\boldsymbol{k}}(0)\delta\rho_{\boldsymbol{k}}(0)\rangle} 
= \frac{P(k,s)}{Q(k,s)} , \label{autocorrdens}
\end{equation}
where
\begin{subequations}
\begin{eqnarray}
  P(k,s) & = & \left(s+ \gamma_0 \chi_0 k^2\right) \left(s+\nu_{l0} k^2\right) 
  \left(s+ \eta_0 k^2\right) \left(s+ \nu_{s0} k^2\right) \, \; \nonumber \\
  && \, + \frac{\gamma_0-1}{\gamma_0} k^2 c_{s0}^2  \left(s+ \eta_0 k^2\right) \left(s+ \nu_{s0} k^2\right) 
  \nonumber \\
 && \, + k^2 v_A^2 \left(s+ \gamma_0 \chi_0 k^2\right) \left(s+k^2 \left[ \nu_{s0}  + \nu_{c0} \cos^2{\theta} \right] \right) 
  \nonumber \\
  && + \frac{\gamma_0-1}{\gamma_0} k^4 v_A^2 c_0^2 \cos^2{\theta} , \\
  Q(k,s) & = & \frac{k^2 c_{s0}^2}{\gamma_0} \left(s+ \gamma_0 \chi_0 k^2\right)
  \left(s+ \eta_0 k^2\right) \left(s+ \nu_{s0} k^2\right) \nonumber \\
  && \, + \frac{1}{\gamma_0} k^4 v_A^2 c_0^2 \cos^2{\theta} \left(s+ \gamma_0 \chi_0 k^2\right) 
  + s P(k,s) .\quad \label{specialfuncs}
\end{eqnarray}
\end{subequations}
Here, we define various additional quantities: shear kinematic viscosity $\nu_{s0} \equiv \zeta_{s0}/\rho_0$, compressive 
kinematic viscosity $\nu_{c0} = \zeta_{c0}/\rho_0$, longitudinal viscosity $\nu_{l0} = \nu_{s0} + 
\nu_{c0}$, $\theta$ the angle between $\boldsymbol{B}_0$ and $\boldsymbol{k}$, and $v_A$ the Alfv\`en speed:
\begin{equation}
  v_A \equiv \frac{B_0}{\sqrt{\mu_0 \rho_0}} \, ,
\end{equation}
(where $B_0 = |\boldsymbol{B}_0|$).
Full details of this calculation are presented in 
Appendix \ref{AppendixDensity}; we note for clarity's sake that the derivation  of 
(\ref{autocorrdens}) assumes that the initial density fluctuations are 
uncorrelated with the initial velocity, magnetic field and temperature 
fluctuations. 

In principle, one can now calculate the dynamic structure factor using 
(\ref{DSFlimit}); however, this is algebraically tedious, and the general result initially rather 
opaque. It is more physically elucidating to instead consider various special 
cases of the dynamic structure factor, where analytical calculations can be undertaken more readily.
These are presented next. 

\section{Parallel fluctuations}
\label{Fluctuations_par}

We first consider fluctuations whose wavevector is parallel to the magnetic field: 
in other words, $\cos{\theta} = 1$.  In this case, it is elementary to show that
\begin{subequations}
\begin{eqnarray}
P(k,s) & = & \left[\left(s+ \eta_0 k^2\right) \left(s+ \nu_{s0} k^2\right) + k^2 v_A^2\right] P_{\|}(k,s) , \qquad \; \\
Q(k,s) & = & \left[\left(s+ \eta_0 k^2\right) \left(s+ \nu_{s0} k^2\right) + k^2 v_A^2\right] Q_{\|}(k,s) , \qquad \;
\end{eqnarray}
\end{subequations}
where 
\begin{subequations}
\begin{eqnarray}
  P_{\|}(k,s) & = & \left(s+ \gamma_0 \chi_0 k^2\right) \left(s+\nu_{l0} k^2\right) + \frac{\gamma_0-1}{\gamma_0} k^2 c_{s0}^2 , \qquad \\
  Q_{\|}(k,s) & = & \frac{k^2 c_{s0}^2}{\gamma_0} \left(s+ \gamma_0 \chi_0 k^2\right)+ s P_{\|}(k,s) . \;
\end{eqnarray}
\end{subequations}
The density autocorrelation function becomes
\begin{equation}
\frac{\langle\delta \rho_{\boldsymbol{k}}^*(0)\widetilde{\delta\rho}_{\boldsymbol{k}}(s)\rangle}{ \langle \delta \rho^*_{\boldsymbol{k}}(0)\delta\rho_{\boldsymbol{k}}(0)\rangle} 
= \frac{P_{\|}(k,s)}{Q_{\|}(k,s)} . \label{autocorrdens_par}
\end{equation}
We then consider the limit where the dissipation rate of 
fluctuations is much smaller than the frequency, that is
\begin{equation}
|s| \sim \omega \gg \chi_0 k^2, \eta_0 k^2, \nu_{s0} k^2, \nu_{l0} k^2.  \label{weakdampingassum}
\end{equation}
Expanding (\ref{autocorrdens_par}) in this limit (see Appendix \ref{AppendixDSF} for an outline of the expansion technique), and then calculating $S_{nn}(k,\omega)$
using (\ref{DSFlimit}), we find
\begin{widetext}
\begin{equation}
\label{hydrofrac}
{S_{nn}(k, \omega) \over S_{nn}(k)} \approx {\gamma_0-1 \over \gamma_0}{2\chi_0 k^2 \over \omega^2 +  \left(\chi_0 k^2 \right)^2}
+ {1 \over \gamma_0}\left[{\Gamma_{\|} k^2 \over \left(\Gamma_{\|} k^2\right)^2 + \left(\omega +c_{s0}k\right)^2}
\right. + \left. {\Gamma_{\|} k^2 \over \left(\Gamma_{\|} k^2\right)^2 + \left(\omega -c_{s0}k\right)^2}\right],
\end{equation}
\end{widetext}
where
\begin{equation}
\Gamma_{\|} = \frac{(\gamma_0 - 1) \chi_0 + \nu_{l0}}{2}.
\end{equation}

The resulting structure factor is identical to the hydrodynamic structure factor \citep{hansen2}: there are three peaks, two of which are associated with sound waves (so-called Brillouin 
peaks), and one associated with the entropy mode (the elastic peak). 
The location of the Brillouin peaks is given
by the dispersion relation of sound waves; the dependence of their width on both 
viscosity and thermal diffusivity via $\Gamma_{\|}$ is reflective of the fact that both viscous and conductive losses 
damp sound waves. The elastic peak has zero frequency on account of the 
non-propagating nature of the entropy mode; the thermal diffusivity alone 
determines the width, because for small-amplitude fluctuations, conductive 
losses consititute the primary damping mechanism for the mode. 
The reason that the MHD structure factor for parallel wavenumbers is identical to the hydrodynamic one 
is simply that parallel compressive fluctuations in 
MHD do not interact the magnetic field. Parallel fluctuations of the magnetic 
field can exist (in particular, Alfv\`en waves), but do not have a density 
perturbation associated with them. 

\section{Quasi-perpendicular fluctuations}
\label{Fluctuations_perp}

Next, we turn to perturbations which are almost perpendicular to the 
magnetic field; in other words, $\cos{\theta} \ll 1$. We also assume that the thermal and magnetic energy densities 
of the equilibrium are comparable; mathematically, this is equivalent to 
ordering $v_A \sim c_{s0}$. 
This regime is relevant for intergalactic plasma, where tiny magnetic fields are amplified and brought to equipartition via the turbulent dynamo mechanism \citep{kazantsev}. This also applies to laboratory turbulent plasmas when the resistivity is small enough that magnetic field dissipation becomes very weak~\citep{tzeferacos}. 

In this case, again considering the approximation (\ref{weakdampingassum}) it can be shown that (see Appendix \ref{AppendixDSF})
\begin{widetext}
\begin{eqnarray}
\label{quasiperpfrac}
{S_{nn}(k, \omega) \over S_{nn}(k)} & \approx & {\gamma_0-1 \over \gamma_0}{2\chi_0 k^2 \over \omega^2 +  \left(\chi_0 k^2 \right)^2}
+ {1 \over \gamma_0} \frac{v_A^2}{c_{FW}^2} \left[{\Gamma_{SW} k^2 \over \left(\Gamma_{SW} k^2\right)^2 + \left(\omega +c_{SW} k\right)^2}
\right. + \left. {\Gamma_{SW} k^2 \over \left(\Gamma_{SW} k^2\right)^2 + \left(\omega -c_{SW} k\right)^2}\right] \nonumber \\
&& \qquad + {1 \over \gamma_0} \frac{c_{s0}^2}{c_{FW}^2} \left[{\Gamma_{FW} k^2 \over \left(\Gamma_{FW} k^2\right)^2 + \left(\omega +c_{FW}k\right)^2}
\right. + \left. {\Gamma_{FW} k^2 \over \left(\Gamma_{FW} k^2\right)^2 + \left(\omega -c_{FW}k\right)^2}\right],
\end{eqnarray}
\end{widetext}
where
\begin{subequations}
\begin{eqnarray}
c_{SW} & = & \frac{v_A \cos{\theta}}{\sqrt{1+v_A^2/c_{s0}^2}} , \\
c_{FW} & = & \sqrt{c_{s0}^2+v_A^2} ,
\end{eqnarray}
\end{subequations}
and,
\begin{subequations}
\begin{eqnarray}
\Gamma_{SW} & = & \frac{1}{2} \left[\left(\gamma_0-1\right)\frac{v_A^2}{c_{FW}^2} \chi_0 + \frac{c_{s0}^2}{c_{FW}^2} \eta_0 + \nu_{s0} \right], \label{quasiperp_diffuse_SW} \\
\Gamma_{FW} & = & \frac{1}{2} \left[\left(\gamma_0-1\right) \frac{c_{s0}^2}{c_{FW}^2} \chi_0 +  \frac{v_A^2}{c_{FW}^2} \eta_0 + \nu_{l0} \right]. \label{quasiperp_diffuse_FW} 
\end{eqnarray}
\end{subequations}

By comparison to the pure hydrodynamic case (\ref{hydrofrac}), we immediately note a number of 
similarities and differences. Most significantly, the dynamic structure factor for 
quasi-perpendicular modes has five peaks rather than three. The elastic peak 
remains unchanged, and there still exist two peaks at frequencies $\omega \gtrsim k 
c_{s0}$. However, the frequency position of these peaks is now also dependent on the magnetic 
field, and is greater than the sound speed: $\omega = k c_{FW} > k c_{s0}$. 
In addition, a new pair of peaks has emerged with characteristic frequency much 
smaller than the sound speed ($\omega = k c_{SW} \sim k v_A \cos{\theta} \ll k c_{s0}$). 
The width and heights of both peaks are comparable, and depend on the 
viscosity, resistivity and thermal diffusivity. 

The emergence of the additional peaks and their subsequent characteristics can again be explained physically. More 
specifically, in MHD one finds two distinct quasi-perpendicular modes with density perturbations: 
the fast and slow magnetosonic waves. Fast magnetosonic waves are conceptually similar to 
sound waves, except for the effective equilibrium pressure being increased by 
additional magnetic pressure: in fast waves, the magnetic and thermal pressure fluctuations are in phase. 
By constrast, slow magnetosonic waves are almost incompressible ($\nabla \cdot \delta \boldsymbol{u} \sim \cos{\theta} \ll 1 
$), with magnetic and thermal pressure fluctuations acting out of phase. Both magnetosonic waves in general have 
significant magnetic and thermal components, and thus are both subject to 
resistive and conductive damping; however, the effective viscosity experienced 
by the waves is different, on account of the quasi-incompressibility of slow
magnetosonic waves. Finally, the entropy mode is unchanged in MHD, and does not 
have a magnetic component; thus, it is not surprising that the 
elastic peak is unchanged.

\section{Oblique fluctuations}
\label{Fluctuations_oblique}

Finally considering the case of oblique fluctuations, we can now anticipate that the 
dynamic structure factor will have five peaks: an entropy peak, and four 
additional peaks. The approximate positions of these peaks can be obtained by 
considering roots of $Q(k,s)$ when all diffusive effects are completely 
neglected:
\begin{equation}
Q(k,s) \approx s  \left[s^4 + \left(k^2 c_{s0}^2 + k^2 v_A^2\right)s^2 + k^4 v_A^2 c_{s0}^2 \cos^2{\theta}\right] 
\, .
\end{equation}
The five roots are then
\begin{equation}
s^2 = 0, - \frac{1}{2} k^2 \left[c_{s0}^2 + v_A^2 \pm \sqrt{\left(c_{s0}^2 + v_A^2\right)^2 - 4 c_{s0}^2 v_A^2 \cos^2{\theta}}\right]  
\, ,
\end{equation}
with associated peak frequencies
\begin{equation}
\omega^2 = 0,  \frac{1}{2} k^2 \left[c_{s0}^2 + v_A^2 \pm \sqrt{\left(c_{s0}^2 + v_A^2\right)^2 - 4 c_{s0}^2 v_A^2 \cos^2{\theta}}\right]  
\, .
\end{equation}
The `+' roots correspond to the fast magnetosonic modes, and the `$-$' roots to 
the slow magnetosonic modes. We note that for $c_{s0} \sim v_A$, and $\cos{\theta} \lesssim 
1$, the fast and slow magnetosonic modes have comparable frequencies, but that the fast magnetosonic mode's frequency is always 
greater. For quasi-parallel modes, the frequencies are very similar; for 
quasi-perpendicular, they have different orders of magnitude. 
The width and height of the peaks are controlled by (in general, quite complicated) linear combinations 
of the resistivity, thermal diffusivity and the viscosities. 

\begin{figure*}
  \centering
\includegraphics[width=0.65\linewidth]{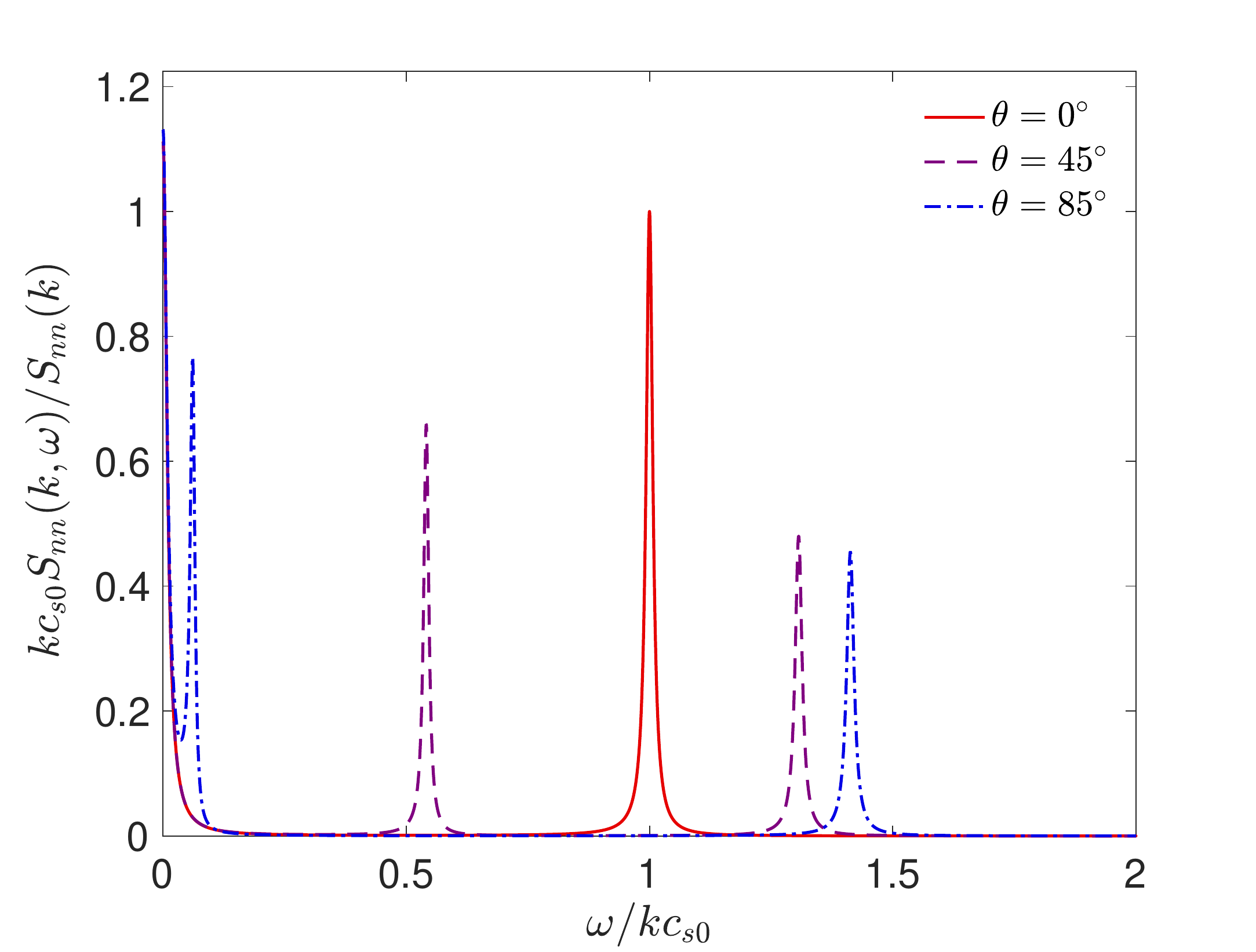}
\caption{\textit{The dynamic structure factor in magnetized, high density plasma}. At a given angle, the dynamic structure factor is calculated by first evaluating the density
autocorrelation function using (\ref{autocorrdens}), before taking the limit 
specified in (\ref{DSFlimit}). The structure factor is presented in a  
dimensionless form; this is obtained via
$s \mapsto s k c_{s0}$.  With this mapping, the magnitude of the various dissipative terms are represented by 
the dimensionless numbers $k \chi_0/c_{s0}$, $k \eta_0/c_{s0}$, $k \nu_{s0}/c_{s0}$ and $k 
\nu_{c0}/c_{s0}$. Three angles are plotted in Figure 1: parallel ($\theta = 0^{\circ}$), 
oblique ($\theta = 45^{\circ}$) and quasiparallel ($\theta = 85^{\circ}$). The 
peak magnitude in each example is normalised to the parallel case. For this particular plot, 
we choose $v_A = c_{s0}$, $k \chi_0/c_{s0} = 0.01$, $k \eta_0/c_{s0} = 0.01$, $k \nu_{s0}/c_{s0} = 0.0067$ and $k 
\nu_{c0}/c_{s0} = 0.0033$.} \label{fig:DSFangle}
\end{figure*}

These claims are illustrated in Figure \ref{fig:DSFangle}, which shows the 
dynamic structure factor evaluated numerically using equations (\ref{DSFlimit}) and (\ref{autocorrdens})
for $v_A = c_{s0}$, and weak dissipation terms (see caption for details). 
As anticipated, for parallel modes the MHD dynamic structure factor is identical 
to the hydrodynamic one. However, for oblique modes, we observe two scattering 
peaks at positive frequencies, whose positions become strongly separated when 
fluctuations close to perpendicular are considered.

\begin{figure*}
  \centering
\includegraphics[width=0.65\linewidth]{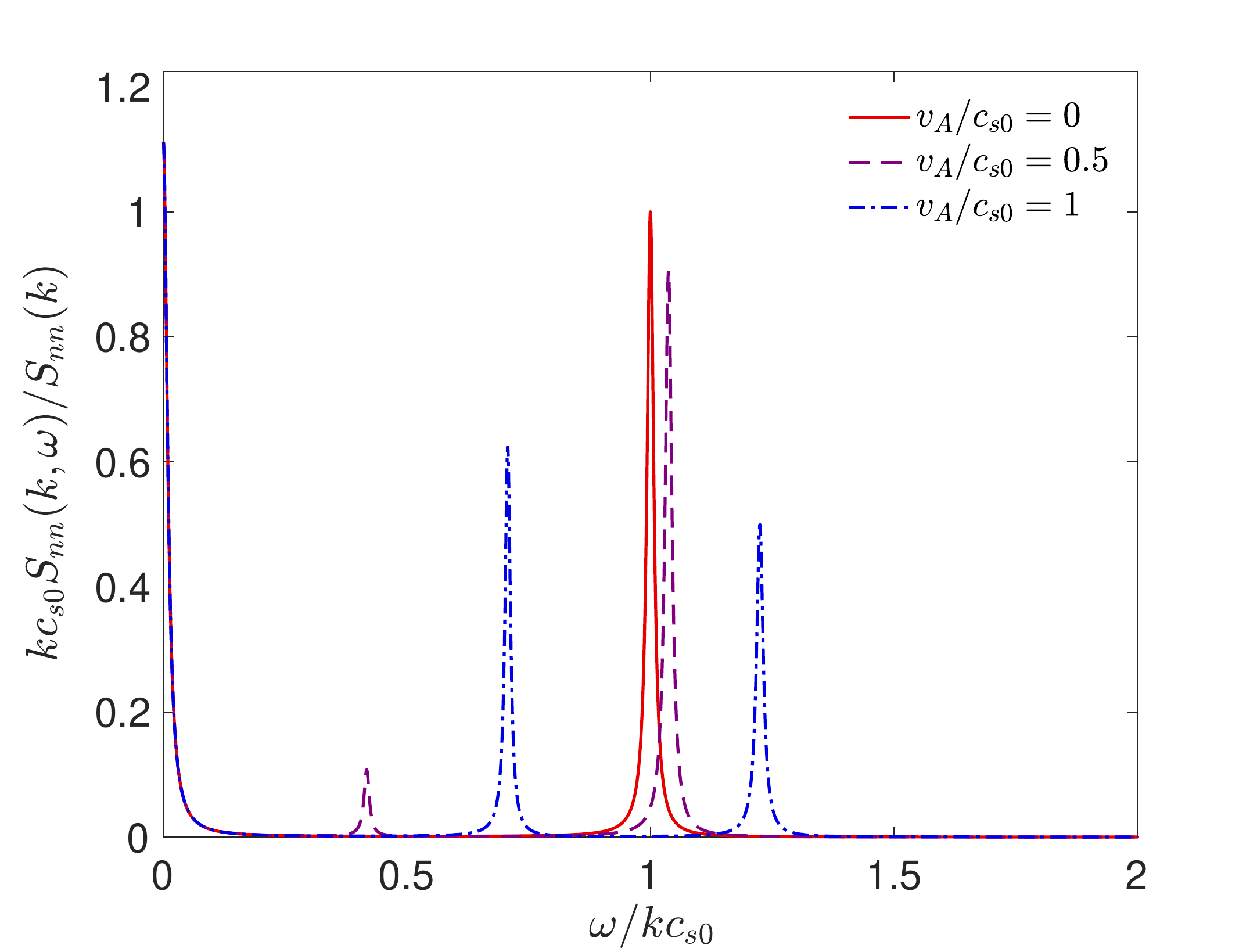}
\caption{\textit{The dynamic structure factor in magnetized, high density plasma with increasing magnetization}. The plotted dynamic
structure factors are calculated in the same way as Figure \ref{fig:DSFangle}, with the same dimensionless values for the dissipative terms. However, in contrast 
to Figure \ref{fig:DSFangle}, the angle of fluctuations with respect to the magnetic field $\theta$ is fixed at $\theta = 30^{\circ}$, while the strength of the magnetic field is increased from nothing ($v_A/c_{s0} = 0$) via a subdominant magnetic field ($v_A/c_{s0} = 0.5$) to a field in equipartition ($v_A/c_{s0} = 1$).} 
\label{fig:DSFweakfield}
\end{figure*}

One further special case which can be treated analytically is that of a weak, but finite 
magnetic field: $v_A \ll c_{s0}$, but $v_A/k \chi_0, v_A/k \eta_0, v_A/k \nu_{l0} \gg 
1$. As well as being tractable, this regime is conceptually interesting because 
it is relevant to understanding the transition between unmagnetized and 
magnetized matter. In this regime, the frequency $\omega_{FW}$ of the fast magnetosonic mode greatly exceeds 
the slow mode $\omega_{SW}$:
\begin{eqnarray}
\omega_{FW} & \approx & k c_{s0} \gg \omega_{SW} \approx k v_A \cos{\theta} . 
\end{eqnarray} 
This separation of frequencies again allows for an analytical form of the 
dynamic structure factor to be derived:
\begin{widetext}
\begin{eqnarray}
\label{weakfieldfrac}
{S_{nn}(k, \omega) \over S_{nn}(k)} & \approx & {\gamma_0-1 \over \gamma_0}{2\chi_0 k^2 \over \omega^2 +  \left(\chi_0 k^2 \right)^2}
+ {1 \over \gamma_0} \frac{v_A^2}{c_{s0}^2} \left[{\Gamma_{A} k^2 \over \left(\Gamma_{A} k^2\right)^2 + \left(\omega +c_{A} k\right)^2}
\right. + \left. {\Gamma_{A} k^2 \over \left(\Gamma_{A} k^2\right)^2 + \left(\omega -c_{A} k\right)^2}\right] \nonumber \\
&& \qquad + {1 \over \gamma_0}\left[{\Gamma_{\|} k^2 \over \left(\Gamma_{\|} k^2\right)^2 + \left(\omega +c_{s0}k\right)^2}
\right. + \left. {\Gamma_{\|} k^2 \over \left(\Gamma_{\|} k^2\right)^2 + \left(\omega -c_{s0}k\right)^2}\right],
\end{eqnarray}
\end{widetext}
where
\begin{subequations}
\begin{eqnarray}
c_{A} & \equiv & v_A \cos{\theta} , \\
\Gamma_{A} & = & \frac{ \eta_0 + \nu_{s0}}{2}  . 
\end{eqnarray}
\end{subequations}
In this regime, we see that both the Brillouin peaks and the entropy peak
remain unaltered from their hydrodynamic form; however, an additional peak 
exists, whose peak amplitude is proportional to $v_A^2/c_{s0}^2$. Thus, as the magnetisation increases, it is anticipated that an additional pair of peaks would emerge in the dynamic structure factor, 
with their amplitude solely a function of the magnetic field strength. In 
addition, we note that the width of these additional peaks is a function of the 
resisitvity and kinematic shear viscosity alone. This is because slow 
magnetosonic waves in the limit $v_A \ll c_{s0}$ have asymptotically small temperature and compressive 
velocity perturbations, and so dissipation via the bulk viscosity or thermal 
diffusivity is very weak. These claims are demonstrated numerically in Figure 
\ref{fig:DSFweakfield}, where for a fixed angle increasing values of $v_A/c_{s0}$ 
are presented.

\section{Detailed balance}
\label{balance}
So far we have discussed the scattering cross section as essentially a classical process. We expect this to be a good approximation since we are dealing with the ion dynamics \citep{gregori2}. Higher order quantum corrections associated to diffraction and nonlocality can, in principle, be included in the MHD formalism discussed here via the introduction of the Bohm potential \citep{schmidt,cross2}, while other effects are implicitly accounted for from the specific form of the transport coefficients.  

On the other hand, quantum effects directly associated with detailed balance are not always negligible, especially if we are dealing with low frequency excitations, as in the present work.
These effects, however, can be brought back in the cross section via the prescription
\citep{scopigno,gregori2}
\begin{equation}
    S_{nn}(\boldsymbol{k},\omega) \rightarrow \frac{\hbar \omega/k_B T}{1-e^{-\hbar \omega/k_B T}} S_{nn}(\boldsymbol{k},\omega).
\end{equation}

\section{Concluding remarks}
In this paper we have discussed the structure of the Thomson scattering cross section in a non-relativistic, dense, magnetized plasma, where collective excitations are most appropriately described via magnetohydrodynamics. We have found that, in addition to cyclotron resonances, the form of the structure factor is dependent on the angle of fluctuations 
with respect to the large-scale magnetic field present in the matter. For 
parallel fluctuations, the dynamic structure factor is the same as the 
hydrodynamic one. However, for oblique fluctuations an additional pair of peaks 
emerges, which are associated with fluctuations of the magnetic field in combination with 
density fluctuations. For quasi-perpendicular fluctuations, there exists a large discrepancy between 
the frequency of the two peaks, of the order of the parallel wavenumber divided 
by the total wavenumber. The existence of the additional pair of peaks holds irrespective of the exact nature of momentum and heat transport in the plasma, provided the general diffusion rates associated with the transport are small compared to the frequencies of fluctuations.

In addition, we observe that the qualitative features of the calculated Thomson scattering cross section also apply to all collisional magnetized plasmas, strongly coupled or not, provided frequencies $\omega$ and wavenumbers $k$ are sufficiently small when compared to electron collision rates and mean-free-paths. This is because magnetohydrodynamics is an appropriate model for weakly coupled collisional plasma on large scales -- indeed, exact constitutive relations for such plasma can be derived formally using kinetic theory~\citep{braginskii}. We note that the Thomson scattering cross section has been evaluated previously for weakly coupled plasma using kinetic theory, including a collision operator~\citep{froula}. However, such calculations usually assume that the equilibrium distribution of the plasma is Maxwellian, with spatially constant macroscopic parameters (density, temperature and magnetic field). By contrast, large-scale magnetohydrodynamic modes require spatially varying macroscopic parameters, and so the associated peaks do not seem to be captured in this previous work. That being said, it should be emphasized that the kinetic theory calculations describe collective excitations in a weakly coupled plasma on small wavenumber scales $k \lambda_{\rm mfp} \gtrsim 1$ which are not present in the magnetohydrodynamical model.

 
From an experimental point of view, we note that the existence of many peaks in the scattering spectra 
allow for the simultaneous measurement of the sound speed and the magnetic 
field in magnetized dense plasma. Furthermore, the width of both peaks 
can be used to constrain transport properties in the plasma, although the resistivity, bulk 
viscosity and shear viscosity cannot be measured simultaneously (unless one or more of these transport 
coefficients is known to be small). This implies that Thomson scattering can be implemented as a powerful diagnostics tool for plasma properties that are otherwise very challenging to measure \citep{evans}.

		

\acknowledgements
The research leading to these results has received funding from
AWE plc., and the Engineering and Physical Sciences Research Council (grant numbers EP/M022331/1 and EP/N014472/1) of the United Kingdom.

\FloatBarrier

%

\appendix

\vspace{12pt}

\section{A: Thermodynamic identities}
\label{AppendixThermo}

In this appendix, we derive equations (\ref{thermoiden_A}) and (\ref{thermoiden_B}) for the specific entropy and pressure in terms of state variables temperature and density and constitutive parameters of the matter. 
Assuming that specific entropy $S = S(\rho,T)$, the total differential is given by
\begin{equation}
    \mathrm{d}S = \left(\frac{\partial S}{\partial \rho}\right)_{T} \mathrm{d} \rho + \left(\frac{\partial S}{\partial T}\right)_{\rho} \mathrm{d} T .
\end{equation}
Using the reciprocal identity, reciprocity and Maxwell's identities, it follows that
\begin{equation}
\left({\partial S \over \partial \rho} \right)_T = -{1 \over \rho^2}\left({\partial p \over \partial T}\right)_\rho = {C_{V} - C_{P} \over \alpha_{T} \rho T},
\quad \left({\partial S \over \partial T}\right)_\rho = {C_{V} \over T},
\end{equation}
where $C_P$ is the heat capacity at constant pressure, and the coefficient of thermal expansion is defined by $\alpha_T = - \rho^{-1} (\partial \rho/\partial T)_{p}$. We conclude that
\begin{equation}
    \mathrm{d}S = \frac{C_V}{T}  \left(\mathrm{d} T 
 - \frac{\gamma - 1}{\alpha_T} \mathrm{d} \rho \right),
\end{equation}
where we have used $\gamma = C_{P}/C_{V}$. This translates immediately into 
(\ref{thermoiden_A}).

Similarly, the pressure $p = p(\rho,T)$ leads to total differential
\begin{equation}
    \mathrm{d}p = \left(\frac{\partial p}{\partial \rho}\right)_{T} \mathrm{d} \rho + \left(\frac{\partial p}{\partial T}\right)_{\rho} \mathrm{d} T .
\end{equation}
Reciprocity and Maxwell's identities then give
\begin{equation}
 {\left({\partial p \over \partial \rho}\right)}_T = {C_{V} \over C_{P}}{\left({\partial p \over \partial \rho}\right)}_{S} = {c_{s}^2 \over \gamma}, 
\quad {\left({\partial p \over \partial T}\right)}_{\rho} = {\left({\partial p \over \partial \rho}\right)}_{T}{\left({\partial \rho \over \partial T}\right)}_{p} = {c_{s}^2 \over \gamma}\alpha_{T} \rho,
\end{equation}
using the definition $c_{s}^2 \equiv (\partial p/\partial \rho)_{S}$ for the adiabatic sound speed. This implies that
\begin{equation}
    \mathrm{d} p = \frac{c_{s}^2}{\gamma} \left(\mathrm{d} \rho + \rho \alpha_T \, \mathrm{d} T\right), 
\end{equation}
from which (\ref{thermoiden_B}) follows trivially. 

\section{B: Solving for the density autocorrelation function}
\label{AppendixDensity}

Here we describe the method used to derive equation (\ref{autocorrdens}) from system of equations (\ref{MHDlintranseqns}).
We begin by assuming that the initial density fluctuations ${\delta\rho}_{\boldsymbol{k}}(0)$ are uncorrelated with the initial temperature fluctuations ${\delta T}_{\boldsymbol{k}}(0)$, the initial velocity fluctuations ${\delta \boldsymbol{u}}_{\boldsymbol{k}}(0)$, and the initial magnetic field fluctuations ${\delta \boldsymbol{B}}_{\boldsymbol{k}}(0)$. This assumption allows for these latter three quantities to be set to zero in (\ref{MHDlintranseqns}b), (\ref{MHDlintranseqns}c) and (\ref{MHDlintranseqns}d) when deriving (\ref{autocorrdens}) without altering the final result.

Next, we write the magnetic field fluctuations $\widetilde{\delta \boldsymbol{B}}_{\boldsymbol{k}}(s)$ and the temperature fluctuations $\widetilde{\delta T}_{\boldsymbol{k}}(s)$ in terms of velocity field fluctuations $\widetilde{\delta \boldsymbol{u}}_{\boldsymbol{k}}(s)$, using (\ref{MHDlintranseqns}c) and (\ref{MHDlintranseqns}d) respectively:
\begin{subequations}
 \label{MHDlintranseqns_solveforBT}
\begin{eqnarray}
\widetilde{\delta \boldsymbol{B}}_{\boldsymbol{k}}(s)  & = & \frac{i \left(\boldsymbol{k} \cdot \boldsymbol{B}_0 \right) \widetilde{\delta \boldsymbol{u}}_{\boldsymbol{k}}(s) - i (\boldsymbol{k} \cdot \widetilde{\delta \boldsymbol{u}}_{\boldsymbol{k}}(s)) \boldsymbol{B}_0}{s+ \eta_0 k^2}, \\
\widetilde{\delta T}_{\boldsymbol{k}}(s) & = & - \frac{i \left(\gamma_0 - 1 \right) \boldsymbol{k} \cdot \widetilde{\delta \boldsymbol{u}}_{\boldsymbol{k}}(s)}{\alpha_{T0}\left(s + \gamma_0 \chi_0 k^2 \right)} .    
\end{eqnarray} 
\end{subequations}

We then solve for $\widetilde{\delta \boldsymbol{u}}_{\boldsymbol{k}}(s) \cdot \boldsymbol{B}_0$ in terms of $\widetilde{\delta\rho}_{\boldsymbol{k}}(s)$ and $i \rho_0 \boldsymbol{k}\cdot \widetilde{\delta \boldsymbol{u}}_{\boldsymbol{k}}(s)$, by taking the scalar product of (\ref{MHDlintranseqns}b) with $\boldsymbol{B}_0$, and substituting (\ref{MHDlintranseqns_solveforBT}a) and (\ref{MHDlintranseqns_solveforBT}b). This gives
\begin{eqnarray}
\widetilde{\delta \boldsymbol{u}}_{\boldsymbol{k}}(s) \cdot \boldsymbol{B}_0 = - \frac{i c_{s0}^2 \left(\boldsymbol{k} \cdot \boldsymbol{B}_0 \right)}{\mu_0 \gamma_{0}\left(\rho_ 0 s + \zeta_{s0} k^2 \right)} \widetilde{\delta\rho}_{\boldsymbol{k}}(s) + \frac{i \left(\boldsymbol{k} \cdot \boldsymbol{B}_0 \right)}{\mu_0 \left(\rho_ 0 s + \zeta_{s0} k^2\right)} \left[\frac{\left(\gamma_0 - 1 \right) c_{s0}^2}{\gamma_{0}\left(s + \gamma_0 \chi_0 k^2 \right)} + \nu_{c0}\right] i \rho_0 \boldsymbol{k}\cdot \widetilde{\delta \boldsymbol{u}}_{\boldsymbol{k}}(s). \label{MHDlintranseqns_solveforuB_0}
\end{eqnarray}
We can subsequently evaluate $i \rho_0 \boldsymbol{k}\cdot \widetilde{\delta \boldsymbol{u}}_{\boldsymbol{k}}(s)$ in terms of $\widetilde{\delta\rho}_{\boldsymbol{k}}(s)$ alone, using the scalar product (\ref{MHDlintranseqns}b) and $i \rho_0 \boldsymbol{k}$, as well as substituting in (\ref{MHDlintranseqns_solveforBT}a) (\ref{MHDlintranseqns_solveforBT}b), and (\ref{MHDlintranseqns_solveforuB_0}):
\begin{eqnarray}
i s \rho_0 \boldsymbol{k} \cdot \widetilde{\delta \boldsymbol{u}}_{\boldsymbol{k}}(s) & = & \left[\frac{k^2 c_{s0}^2}{\gamma_{0}} +\frac{k^2 c_{s0}^2 \left(\boldsymbol{k} \cdot \boldsymbol{B}_0 \right)^2}{\mu_0 \gamma_0 \left(\rho_ 0 s + \zeta_{s0} k^2 \right) \left(s + \gamma_0 \chi_0 k^2 \right)}\right] \widetilde{\delta\rho}_{\boldsymbol{k}}(s) \nonumber \\
&& - \Bigg[\frac{\left(\gamma_0 - 1 \right) k^4 c_{s0}^2}{\gamma_{0}\left(s + \gamma_0 \chi_0 k^2 \right)} + \nu_{l0} k^2 + \frac{k^2 B_0^2}{\mu_0 \rho_0 \left(s + \eta_0 k^2\right)} \nonumber \\
&& + \frac{k^2  \left(\boldsymbol{k} \cdot \boldsymbol{B}_0 \right)^2}{\mu_0 \rho_0 \left(s+ \eta_0 k^2\right) \left(\rho_ 0 s + \zeta_{s0} k^2\right)} \left(\frac{\left(\gamma_0 - 1 \right) c_{s0}^2}{\gamma_{0}\left(s + \gamma_0 \chi_0 k^2 \right)} + \nu_{c0}\right)\Bigg] i \rho_0 \boldsymbol{k} \cdot \widetilde{\delta \boldsymbol{u}}_{\boldsymbol{k}}(s) .
\end{eqnarray}
This can be rearranged to give
\begin{equation}
    i \rho_0 \boldsymbol{k}\cdot \widetilde{\delta \boldsymbol{u}}_{\boldsymbol{k}}(s) = \left(\frac{Q(k,s)}{P(k,s)}-s\right) \widetilde{\delta\rho}_{\boldsymbol{k}}(s) , \label{MHDlintranseqns_solveforku}
\end{equation}
where functions $P(k,s)$ and $Q(k,s)$ are defined by equations (\ref{specialfuncs}a) and (\ref{specialfuncs}b) in the main text.
Finally, we substitute (\ref{MHDlintranseqns_solveforku}) into (\ref{MHDlintranseqns}a), and solve for $\widetilde{\delta\rho}_{\boldsymbol{k}}(s)$ in terms of ${\delta\rho}_{\boldsymbol{k}}(0)$:
\begin{equation}
\widetilde{\delta\rho}_{\boldsymbol{k}}(s) = \frac{P(k,s)}{Q(k,s)} {\delta\rho}_{\boldsymbol{k}}(0)    .
\end{equation}

The density autocorrelation function (\ref{autocorrdens}) then follows immediately. 

\section{C: Analytic calculations of the dynamic structure factor}
\label{AppendixDSF}

In this appendix, we outline the technique used to derive analytic expressions for the dynamic structure factor in the limit of weak damping: that is, fluctuations for which assumptions (\ref{weakdampingassum}) apply. In particular, the technique leads to expression (\ref{hydrofrac}) for the dynamic structure factor associated with parallel fluctuations, (\ref{quasiperpfrac}) for quasi-perpendicular fluctuations, and (\ref{weakfieldfrac}) for oblique fluctuations in the presence of a small but finite magnetic field. 

The technique in general proceeds as follows. First, following assumptions (\ref{weakdampingassum}), we neglect all diffusive effects, and then determine the (imaginary) values $s_{*} = i \omega_{*}$ of $s$ for which density autocorrelation function (\ref{autocorrdens}) vanishes. Equivalently, these values $s$ are the roots of polynomial $Q(k,s)$, for fixed $k$. Then, for each $s_{*}$, we calculate the density autocorrelation function for $s$ in the neighbourhood of $s_{*}$ when diffusive effects are included -- that is, $|s-s_{*}| \sim \chi_0 k^2, \eta_0 k^2, \nu_{s0} k^2, \nu_{l0} k^2 \ll |s_{*}|$. The resulting expression typically possesses the form
\begin{equation}
\frac{\langle\delta \rho_{\boldsymbol{k}}^*(0)\widetilde{\delta\rho}_{\boldsymbol{k}}(s)\rangle}{ \langle \delta \rho^*_{\boldsymbol{k}}(0)\delta\rho_{\boldsymbol{k}}(0)\rangle} 
\approx \frac{\cal{A}_{*}}{s-i\omega_{*} + \Delta \omega_{*}} , \label{autocorrdens_neighbour}
\end{equation}
for $\cal{A}_*$ some characteristic amplitude, and $\Delta \omega_{*} \sim \chi_0 k^2, \eta_0 k^2, \nu_{s0} k^2, \nu_{l0} k^2$ some typical frequency spread. Then applying formula (\ref{DSFlimit}), we conclude that the dynamic structure factor near the peak frequency $\omega_{*}$ is approximately
\begin{equation}
\frac{S_{nn}(k, \omega)}{S_{nn}(k)} \approx \frac{2 \Delta\omega_{*} \cal{A}_{*} }{(\omega-\omega_{*})^2+(\Delta \omega_{*})^2} . \label{structurefactor_analytical}
\end{equation}
The total structure factor is simply the sum over all scattering peak frequencies. 

We illustrate the approach in the case of quasi-perpendicular fluctuations, on account of the novelty of the result. First neglecting all diffusive effects, we find
\begin{equation}
    Q(k,s) \approx s\left(s^2 + k^2 c_{SW}^2\right) \left(s^2 + k^2 c_{FW}^2\right) , \quad P(k,s) \approx -k^2 c_{s0}^2 \left(s^2 +k^2 v_A^2 \cos^2{\theta}\right)/\gamma_0 ,
\end{equation}
where $c_{SW}$ and $c_{FW}$ are given in the main text. The roots are then 
\begin{equation}
    s_{*} = 0, \quad \pm i k c_{SW}, \quad \pm i k c_{FW} .
\end{equation}
We then determine the density autocorrelation function in the neighbourhood of each of these roots in turn. The numerator is $P(k,s_{*}) \neq 0$ in each case.

\begin{itemize}
    \item \underline{$s_{*} = 0$}: let $s = \delta s \sim \chi_0 k^2, \eta_0 k^2, \nu_{s0} k^2, \nu_{l0} k^2$. Then, $Q(k,s) \approx k^4 c_{s0}^2 v_A^2 \cos^2{\theta} \, (\delta s + \chi_0 k^2)$, and so
\begin{equation}
\frac{\langle\delta \rho_{\boldsymbol{k}}^*(0)\widetilde{\delta\rho}_{\boldsymbol{k}}(s)\rangle}{ \langle \delta \rho^*_{\boldsymbol{k}}(0)\delta\rho_{\boldsymbol{k}}(0)\rangle} 
\approx \frac{(\gamma_0-1)/\gamma_0}{s+ \chi_0 k^2} . \label{autocorrdens_neighbour_entropymode}
\end{equation}
\item \underline{$s_{*} = \pm i k c_{SW} $}: let $s = \pm i k c_{SW} + \delta s$, $\delta s \sim \chi_0 k^2, \eta_0 k^2, \nu_{s0} k^2, \nu_{l0} k^2$. Then $Q(k,s) \approx -2 k^2 c_{SW}^2 c_{FW}^2 \, (\delta s+\Gamma_{SW})$, where $\Gamma_{SW}$ is defined by (\ref{quasiperp_diffuse_SW}) in the main text. It follows that
\begin{equation}
\frac{\langle\delta \rho_{\boldsymbol{k}}^*(0)\widetilde{\delta\rho}_{\boldsymbol{k}}(s)\rangle}{ \langle \delta \rho^*_{\boldsymbol{k}}(0)\delta\rho_{\boldsymbol{k}}(0)\rangle} 
\approx \frac{v_A^2/2 \gamma_0 c_{FW}^2}{s \mp i k c_{SW} + \Gamma_{SW}} . \label{autocorrdens_neighbour_slowmode}
\end{equation}
\item \underline{$s_{*} = \pm i k c_{FW} $}: let $s = \pm i k c_{FW} + \delta s$, $\delta s \sim \chi_0 k^2, \eta_0 k^2, \nu_{s0} k^2, \nu_{l0} k^2$. Then $Q(k,s) \approx -2 k^4 c_{FW}^4 \, (\delta s+\Gamma_{FW})$, where $\Gamma_{FW}$ is also defined in the main text, by (\ref{quasiperp_diffuse_FW}). We conclude that
\begin{equation}
\frac{\langle\delta \rho_{\boldsymbol{k}}^*(0)\widetilde{\delta\rho}_{\boldsymbol{k}}(s)\rangle}{ \langle \delta \rho^*_{\boldsymbol{k}}(0)\delta\rho_{\boldsymbol{k}}(0)\rangle} 
\approx \frac{c_{s0}^2/2 \gamma_0 c_{FW}^2}{s \mp i k c_{FW} + \Gamma_{FW}} . \label{autocorrdens_neighbour_fastmode}
\end{equation}
\end{itemize}
The dynamic structure function near each root is then given by (\ref{structurefactor_analytical}), with the total structure factor (\ref{quasiperpfrac}) simply being the sum of each of these terms. 

\end{document}